\documentclass[12pt]{article}
\usepackage{graphicx}
\usepackage{amsmath}
\usepackage{bm}
\def\Vec#1{\mbox{\boldmath $#1$}}

\def\itmb{\begin{itemize}}
\def\itme{\end{itemize}}
\def\enmb{\begin{enumerate}}
\def\enme{\end{enumerate}}
\def\eqnb{\begin{equation}}
\def\eqne{\end{equation}}

\def\NPB{{Nucl. Phys.} { B}}
\def\PLB{{Phys. Lett.} B}
\def\PRL{Phys. Rev. Lett.}
\def\PRD{{Phys. Rev.} D}

\title{The magnetic mass of transverse gluon, the  B-meson weak decay vertex and the triality symmetry of octonion}

\author{Sadataka Furui \\
  Faculty of Science and Engineering, Teikyo University\\
1-1 Toyosatodai, Utsunomiya, 320-8551 Japan {\thanks
{\textit{E-mail address:} furui@umb.teikyo-u.ac.jp}}
}

\begin{document}
\maketitle
\begin{abstract}%
With an assumption that in the Yang-Mills Lagrangian,  a left-handed fermion and a right-handed fermion both expressed as the quaternion make an octonion which possesses the triality symmetry, I calculate the magnetic mass of the transverse self-dual gluon from three  loop diagram, in which a heavy quark pair is created and two self-dual gluons are interchanged. 

The magnetic mass of the transverse gluon depends on the mass of the pair created quarks, and in the case of charmed quark pair creation, the magnetic mass $m_{mag}$ becomes approximately equal to  $T_c$  at $T=T_c\sim 1.14\Lambda_{\overline{MS}}\sim 260$MeV.

A possible time-like magnetic gluon mass from two self-dual gluon exchange is derived, and corrections in the  B-meson weak decay vertices from the two self-dual gluon exchange are also evaluated.
\end{abstract}
\newpage
\section{Introduction}
In 1980, Linde\cite{Linde80} pointed out difficulties of infrared problems in the thermodynamics of mass-less Yang-Mills gas, and a possible solution via gluon acquiring the magnetic mass.  
At zero temperature, an effective gluon mass of $500\pm 200$ MeV was predicted by Cornwall\cite{Corn82}.
A problem that at relatively low temperature or high density, the pressure of the QCD thermodynamic gas becomes negative was pointed out in \cite{Ka84} and difficulties in the calculation of the thermodynamical potential in the $g^6$ order are discussed in\cite{BN95, BN96, IKRV06}.

The ground-state energy of the finite temperature quark-gluon system was classically derived from the Yang-Mills Lagrangian by Freedman and McLarren\cite{FM77}. An extensive review of calculations done before 2004 are given in \cite{KR04}.  

 In QCD, the gluon is screened in the plasma through gluon loops, quark loops and ghost loops.  In 1993 the non Abelian Debye screening mass was found to be sensitive to the non perturbative  magnetic mass of the gluon\cite{Re93}.   
 By adding a magnetic mass $m_{mag}^2$ to the transverse self-energy $\Pi_T(K)$, the instability could be evaded,  when $m_{mag}=c_f \frac{3}{32}g^2 N_c T$ and  $c_f\geq 1$.  $m_{mag}$ is expected to appear in the perturbative calculation in the order of $g^6$ and higher, but no systematic calculation of it was found.

Alexanian and Nair\cite{AN95} calculated the magnetic mass of the gluon in a model based on Chern-Simons theory and obtained $\displaystyle m_{mag}\sim (2.384)\frac{Cg^2T}{4\pi}$, where $C=N$ for SU(N) gauge theory. There is, however, other models that yield different values of the magnetic mass\cite{JP95, Na98}, and the situation is worrisome. The best that one could do perturbatively to get the Debye mass is\cite{AY95}
 \[
m_D=(\frac{N_c}{3}+\frac{N_f}{6})^{1/2}gT+\frac{1}{4\pi}N_c g^2 T log(\frac{N_c}{3g^2}+\frac{N_f}{6g^2})^{1/2})+c_f g^2 T+O(g^3 T)
\]
 where $N_c$ is the number of colors and $N_f$ is the number of fermion flavors, but the value of $c_f$ is left open.
             
There is a systematic investigation of vacuum polarization tensor $\Pi_{\mu\nu}(k_0,k)$ of relativistic plasma\cite{We92}.  The collective plasma effects are characterized by the frequency $\displaystyle\omega_k=\sqrt{\frac{N_f}{6}+\frac{N_c}{3}} gT$. The self-energy depend on four vector $u^\alpha$ of the fluid and the momentum $K^\alpha$ of the virtual particle via two invariants $\omega=K^\alpha u_\alpha$ and $k$ such that $K^\alpha K_\alpha=\omega^2-k^2$. The transverse self-energy function $\Pi_T(k,\omega)$ and the longitudinal self-energy function $\Pi_L(\omega,k)$ are calculated by a pair of four vector $\tilde E^\alpha$ and $\tilde B^\alpha$, each orthogonal to $u^\alpha$, defined by
\[
F^{\alpha\beta}= u^\alpha\tilde E^\beta -u^\beta \tilde E^\alpha+\epsilon^{\alpha\beta\gamma\delta}\tilde B_\gamma u_\delta
\]
When $\omega>\omega_p$, plane wave solution of $\tilde E^\alpha$ and $\tilde B^\alpha$ exist, but  when $\omega<\omega_p$ they are screened. 
It was shown that the screening length for the magnetic fields diverges at $\omega\to 0$, thus
$\tilde B^\alpha$ is screened except at $\omega=0$.

In \cite{SF09,SF10,SF11}, I proposed a calculation of the Domain Wall Fermion(DWF) propagator in quaternion basis and expressed vector fields in terms of the Pl\"ucker coordinate, following \'E. Cartan\cite{Cartan66}. In this framework, the spin structure of the quark pair in the self-energy diagram with two self-dual gluon exchange is uniquely defined when the color and spin of the incoming gluon are fixed. The process is similar to the one investigated in the technicolor theory\cite{CL84}.

In order to calculate $\tilde B^\alpha$ at $\omega=0$, I adopt a model, in which a quark pair is created, and two self-dual gluons are exchanged and the pair is annihilated\cite{SF10,SF11}. From a left-handed quark described by a quaternion and a right-handed quark described by another quaternion, one can construct an octonion. An octonion posesses the triality symmetry\cite{Cartan66, Lou00} and the spin structure of the magnetic coupling of a self-dual gluon to the quark loops can be fixed when the exchanged internal gluons are restricted to be self-dual. There is no direct evidence that the triality symmetry plays a role in the nature, but the difference of about a factor of 3  in the effective flavor number for opening the conformal window in the Schr\"odinger functional scheme\cite{AFN08} v.s. in the momentum subtraction (MOM) scheme  \cite{FN07b} could be understood,  if the Schr\"odinger functional does not select a triality sector but MOM scheme selects one triality sector.
Using the Banks-Zaks expansion\cite{BZ82}, Grunberg\cite{Gr01} showed that the non-perturbative effect modifies the critical flavor number of the perturbative QCD which was around 8\cite{AFN08}  to 4.
Infrared stable QCD running coupling was observed also in holographic QCD\cite{BTD10},  and in the polarized electron nucleon scattering\cite{JLab07}.

In order to check the relevance of self-dual two gluon exchange in the coupling vertex, I calculate the vertex of a B meson weak decay vertex into lepton and anti-neutrino. 
The decay propability of a B meson into a lepton and neutrino is measured by the Babar collaboration\cite{Babar07} and by the Belle collaboration\cite{Belle10}. Possible deviation from the standard model 
via Penguin diagram is discussed in \cite{LS10} and I compare with the possible correction from two self-dual gluon exchange diagrams.

The structure of this paper is as follows. In the sect.2, I summarize the magentic mass problem and  present  the calculation of the self-energy using the quaternion bases. 
In sect.3, the B-meson weak decay vertex including the two self-dual magnetic gluon exchange
is investigated. Conclusion and discussion are given in the sect.4.

\section{The self-energy of a gluon via exchange of self-dual gluons between heavy quark and heavy anti quark}
In 1993 the non abelian Debye screening mass was found to be sensitive to the non perturbative  magnetic mass of of the gluon\cite{Re93}.   The importance of the magnetic mass was recognized 
in \cite{Ka84}, as he calculated the pressure of finite temperature plasma, which is defined as $p=-T\log Z/V$ where $Z$ is the partition function.  The pressure derived at a high-temperature region
\begin{eqnarray}
p&=&\frac{8}{45}\pi^2 T^4+N_f(\frac{1}{2}T^2\mu^2+\frac{7}{60}\pi^2T^4)\nonumber\\
&&-\frac{16\pi^2}{(22-\frac{4}{3}N_f)\log(T/\Lambda_{QCD})} [\frac{1}{6}T^4+N_f(\frac{5}{72}T^4+T^2\mu^2/4\pi^2)]
\end{eqnarray}
becomes negative when extrapolated to low temperature  $T$ or to high chemical potential $\mu$.
The negative pressure implies that the convergence of the perturbation series breaks down,
 and a QCD phase transition at this point was discussed. 

In the infrared, however, the ring sum term
$log(1+\Pi_T(K)/K^2)$ was found to contain infrared singular term, indepent of the
choice of the gauge.
In order to evade the equation of the one loop transverse gluon propagator near $T=0$
\[
k^2+\Pi_T(k_0=0,k)=k^2-g^2 N_c T\frac{8+(\xi+1)^2}{64}k=0
\]
be satisfied for positive $k$, a modification $\Pi_T(K)\to \Pi_T(K)+m_{mag}^2$ was proposed\cite{IKRV06}. 
With an arbitrary parameter $c_f\geq 1$, $m_{mag}=c_f \frac{3}{32}g^2 N_c T\sim c_f\frac{N_c}{3}T$ was expected  for $\alpha_s(k)=0.3$ at $k\sim 1$GeV.
The magnetic mass contributes to the pressure of the quark-gluon plasma in the order of $g^6T^4$.
In lattice simulations, the magnetic gluon propagator does not show  a peak at the zero momentum and the pole structure of the magnetic mass was, excluded\cite{CKP01}, although how to detect the timelike pole on the lattice remains an open question\cite{Corn09}.

 I do not assume a magnetic mass of $g^2 T$ for a producion of a $g^6$ term in the pressure, but I consider a three loop diagram of order $g^6$ including two self-dual gluon exchange, as a building block of the stabilized quark-gluon plasma system.  The model does not contradict with the lattice simulation.

In \cite{SF09,SF10},  I discussed the quark propagator using quaternion\cite{Lou00} basis. In this flamework,
  I consider tensor coupling of the external gluon field to a heavy quark internal self-dual vector field  $x_1,x_2,x_3$ couple to the fermion spinors. 

\begin{figure}
\begin{minipage}[b]{0.47\linewidth}
\begin{center}
\includegraphics[width=6cm,angle=0,clip]{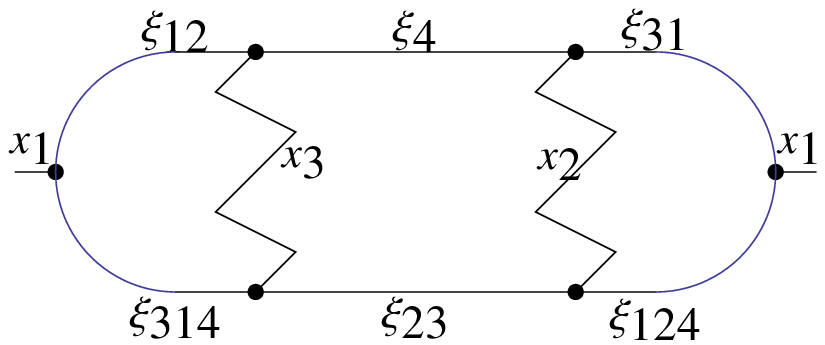}
\caption{The self-energy diagram of transverse polarized gluon through exchange of self-dual gauge fields. 
$\Pi_{11}$}
\label{11a}
\end{center}
\end{minipage}
\hfill
\begin{minipage}[b]{0.47\linewidth}
\begin{center}
\includegraphics[width=6cm,angle=0,clip]{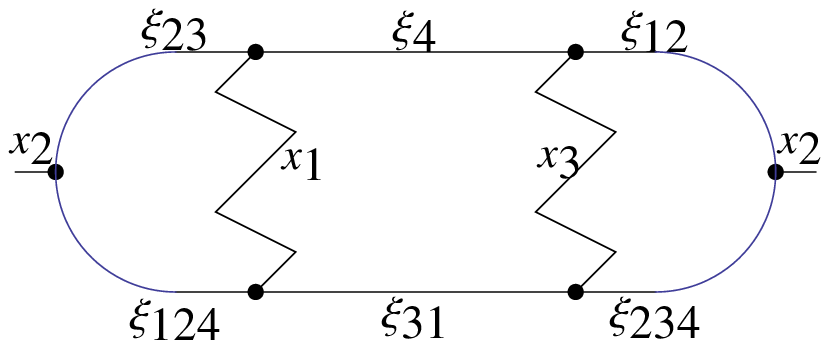}
\caption{The self-energy diagram of transverse polarized gluon through exchange of self-dual gauge fields. $\Pi_{22}$}
\label{g622a}
\end{center}
\end{minipage}
\end{figure}
I consider the self-energy or the magnetic mass of a gluon, which is polarized in the transverse direction. The transverse gluon field is 
\[
F_{\mu\nu}=-ig(\partial_\mu A^a_\nu-\partial_\nu A^a_\mu+g f^{abc}A_\mu^b A_\nu^c)\lambda^a,
\]
where $\lambda^a$ is the SU(3) color basis. The tensor coupling in the case of $\Pi_{T 11}$ are, 
$
\gamma_4(\gamma_2\gamma_3-\gamma_3\gamma_2)/2=\gamma_1\gamma_5
$
in the vertex and in the case of $\Pi_{T 22}$ is
$
\gamma_4(\gamma_3\gamma_1-\gamma_1\gamma_3)/2=\gamma_2\gamma_5
$

The  coupling of $F_{ij}(q)$ to the quark is 
\begin{eqnarray}
&&i\bar \psi(k)\frac{\gamma_i\gamma_j-\gamma_j\gamma_i}{2} \psi(p) F_{ij}(q )\nonumber\\
&&=\psi^*(k)\gamma_5\epsilon_{ij n}\frac{k^j }{m}\sigma^i  \psi(p) (\epsilon_{i j n} p^{i} A_{j} (q)+\cdots)\nonumber\\
&&=\psi^*(k)\gamma_5\epsilon_{ijn}\frac{  p^i  k^j }{m}  \epsilon_{ijn} \sigma^i \psi(p) A_{j}(q)\nonumber\\
&&=\psi^*(k)\gamma_5\frac {(p\times k)_n}{m}  \psi(p) (\sigma\times A(q))_n
\end{eqnarray}

Internal self-dual gluon and the heavy quark couplings in $\Pi_{11a}$  and $\Pi_{22a}$ are

$
\gamma_4(\gamma_3\gamma_4-\gamma_4\gamma_3)/2=-\gamma_3,
$
$
\gamma_4(\gamma_2\gamma_4-\gamma_4\gamma_2)/2=-\gamma_2
$,
and 
$
\gamma_4(\gamma_1\gamma_4-\gamma_1\gamma_3)/2=-\gamma_1.
$

In $\Pi_{11}$, I choose the quark represented by $\xi_4$ to be at rest
and the self-dual gluon $x_3$ and $x_2$ have momenta $k_y$ and $p_z$, respectively.
The quark $\xi_{12}$ and $\xi_{31}$ have momenta $-k$ and $-p$, respectively,
and $\xi_{314}$ and $\xi_{124}$ have momenta $p_z$ and $k_y$, respectively.
The propagator of quark $\xi_{23}$ have the numerator $\gamma_3 p_z+\gamma_2 k_y$, 
but $\gamma_3$ is multiplied by $\gamma_2$ at the junction to the quark $\xi_{314}$,
and $\gamma_2$ is multiplied by $\gamma_3$ at the junction to the quark $\xi_{124}$
and effectively the numerator is proportional to $\sigma_x$, as required by the
assignment of $\xi_{23}$.

I use the Clifford products rule $(a_0+\Vec a)(b_0+\Vec b)=a_0b_0+a_0\Vec b+b_0\Vec a-\Vec a\cdot\Vec b+\Vec a\times\Vec b$ to evaluate Clifford products of the following bases.
\begin{eqnarray}
C\phi&=&\xi_{1234}-\xi_{23}\Vec i-\xi_{31}\Vec j-\xi{12}\Vec k\nonumber\\
C\psi&=&\xi_{123}-\xi_{234}\Vec i-\xi_{314}\Vec j-\xi_{124}\Vec k\nonumber\\
\phi&=&\xi_0+\xi_{14}\Vec i+\xi_{24}\Vec j+\xi{34}\Vec k\nonumber\\
\psi&=&\xi_4+\xi_{1}\Vec i+\xi_{2}\Vec j+\xi_{3}\Vec k
\end{eqnarray}
 
The $x_1$ gluon coupling to quark pair with 2 self-dual gluon exchange $\Pi_{11}^a$ in Coulomb gauge consists of two types i.e  $(-\xi_{12}\xi_{314}+\xi_{31}\xi_{124})$, or $(\xi_{24}\xi_3-\xi_{34}\xi_2)$. The possible quark-anti quark state between the self-dual gluon exchange of the former is $\xi_4\xi_{23}$ 
 and the latter is $\xi_{1234}\xi_1$. 

Since the trace of the two types are the same, I consider the amplitude $\Pi_{11}^a$ of the $x_1$ gluon 
\begin{eqnarray}
&&{J_a(k,p)}=m^4 g^6
\frac{ k_z\times  p_y}{m} Tr\gamma_1\gamma_5\frac{-\gamma_3  k_z+m}{k_z^2+m^2}\gamma_3\frac{1}{k_z^2}\frac{1}{m}\frac{1}{p_y^2}\gamma_2\frac{-\gamma_2 p_y+m }{p_y^2+m^2}\nonumber\\
&&\times \gamma_1\gamma_5 \frac{ k_z\times  p_y}{m} \frac{\gamma_3 k_z+m}{k_z^2+m^2}\gamma_2\gamma_5\frac{\gamma_2 p_y+\gamma_3 k_z+m}{p_y^2+k_z^2+m^2}\gamma_3\gamma_5\frac{\gamma_2 p_y+m}{p_y^2+m^2}
\end{eqnarray}

Similarly, the self-energy from $\Pi_{22}^b$ becomes,  by choosing the intermediate quark-anti quark state $(-\xi_{23}\xi_{124}+\xi_{12}\xi_{234})$, or $(-\xi_{14}\xi_3+\xi_{34}\xi_1)$, 
\begin{eqnarray}
&&{J_b(k,p)}=m^4 g^6
\frac{ k_x\times  p_z}{m} Tr\gamma_2\gamma_5\frac{-\gamma_1 k_x+m }{k_x^2+m^2} \gamma_1\frac{1}{k_x^2}\frac{1}{m}\frac{1}{p_z^2}\gamma_3\frac{-\gamma_3 p_z+m}{ p_z^2+m^2}\nonumber\\
&&\times \gamma_2\gamma_5 \frac{ k_x\times p_z}{m} \frac{\gamma_1 k_x+m}{k_x^2+m^2}\gamma_3\gamma_5\frac {\gamma_3 p_z+\gamma_1 k_x+m}{p_z^2+k_x^2+m^2}\gamma_1\gamma_5\frac{\gamma_3 p_z+m}{p_z^2+m^2} 
\end{eqnarray}

The trace in the numerator of the $\Pi_{11}$
\begin{equation}
4(m^6+3m^4(p_y^2+k_z^2)-m^2p_y^2 k_z^2)
\end{equation}

The numerator of the $\Pi_{22}$ is
\begin{equation}
4(m^6+3m^4(p_z^2+k_x^2)-m^2p_z^2 k_x^2)
\end{equation}

I integrate numerically
\begin{equation}
4m^4\int_0^\infty \int_0^\infty \frac{m^6+3m^4(k^2+p^2)-m^2 k^2 p^2}{(k^2+m^2)^2(p^2+m^2)^2(k^2+p^2+m^2)} \frac{dk dp}{(2\pi)^2} 
\end{equation}
where the factor 4 comes from integral on $k$ and $p$ from $-\infty$ to $+\infty$.

 I chosse the scale $m=1$ and evaluate
\begin{equation}
4\int_0^\Lambda \int_0^\Lambda \frac{1+3(x+y)-x y}{(x+1)^2(y+1)^2(x+y+1)} \frac{dxdy}{(2\pi)^2}\label{integ}
\end{equation}
by varying $\Lambda$. 
Numerically, the integral (\ref{integ}) is approximately ${7}\frac{1}{(2\pi)^2 }$

At $T=0$, when the wave function of the quark pair is available from lattice simulation, one can perform the integration over $q$ numerically and obtain a non-screened magnetic mass. 

 At $T\ne 0$,  it is necessary to replace the self-dual gluon exchange propagator $\displaystyle
\frac{1}{k^2}\frac{1}{p^2}$ in $J(k,p)$ by $\displaystyle \frac{1}{k^2+m_{mag}^2(T)}\frac{1}{p^2+m_{mag}^2(T)}$ and check the self-consistency.  

\vskip 0.2 true cm
In a quenched SU(2) lattice Landau gauge simulation, the temperature dependence of the transverse gluon and longitudinal gluon from 0 temperature to twice the critical temperature $T_c$ were recently 
measured\cite{CM10, CDMV11}.  Near the $T=T_c$, the transverse gluon propagator showed a peak
near $q=0.4$GeV, but the longitudinal gluon propagator did not show this behavior.  In a quenched
SU(3) lattice Landau gauge simulation\cite{NSS04}, the transverse gluon propagator, or the magnetic gluon propagator showed smooth momentum dependence in the range from  $T=0.867 T_c$ to $T=4.97T_c$. 

 In the present work, I ignore these temperatures and the number of color dependence, and
evaluate, using the bose Bose-Einstein distribution for the gluon, the magnetic mass at finite
temperature $T$ as
\begin{equation}
m_{mag}(T)= T\int_0^\infty 4\pi q^2  \frac{f(q)}{e^{q/T}-1} \frac{dq}{(2\pi)^3} 
\end{equation}
Since $q$ is in GeV, I use $k_B=8.6\times 10^{-5}$eV/K$=8.6\times 10^{-14}$GeV/K as the unit, and  express the
temperature in the unit of $T_c=1.14 \Lambda_{\overline{MS}}$ \cite{KR04} and $\Lambda_{\overline{MS}}\sim 230$MeV.

There are two polarization directions and two different time orders, and 
assuming that the coupling of self-dual gluon and the external gluons to quarks are the same, I
estimate the $f(q)$ using the numerical results of $T=0$ as  
$\displaystyle f(q)= g(q)^6\frac{7}{m^3} \frac{4}{(2\pi)^2}=\alpha_s(q)^3\frac{28 \times 16\pi }{m^3}$.
Here, I  adopt $\alpha_s(q)$ obtained from Lattice simulation\cite{FN07b} which is consistent with the prediction of the holographic theory\cite{BTD10}.
The parametrization of $\alpha(s)$ is\cite{JLab07}
\begin{equation}
\alpha_s(q)=\frac{\gamma n(q)}{\log \left(\frac{q^2+m_g(q)^2}{\Lambda^2}\right)}.
\end{equation}
where
\begin{equation}
n(q)=\pi(1+[\frac{\gamma}{\log(m^2/\Lambda^2)(1+q/\Lambda)-\gamma}+(bq)^c]^{-1})
\end{equation}
and $\gamma=\frac{4}{\beta_0}=\frac{12}{33-8}$, $m_g(q)=\frac{m}{1+(aq)^d}$,
 $m=1.024$GeV, $a=3.008$GeV$^{-1}$, $d=0.840$, $b=1.425$GeV$^{-1}$, $c=0.908$ and $\Lambda=0.349$GeV.

\begin{figure}[htb]
\begin{center}
\includegraphics[width=6cm,angle=0,clip]{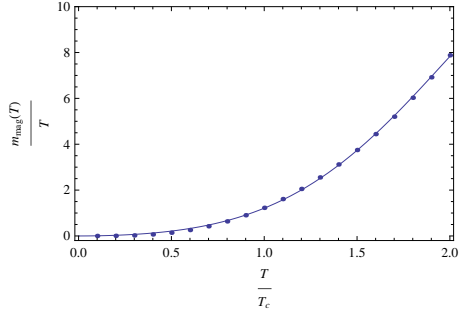}
\caption{The magnetic mass of the gluon through charm quark loops, as a function of $T/T_c$ and a fit.}
\label{magmass_fit}
\end{center}
\end{figure}

The charmed quark-anti quark pair creating 3 loop diagram gives the magnetic mass of the order
of $T_c$ near $T=T_c$.
The magnetic mass  $m_{mag}(T)$in this case can be parameterized as
\[
m_{mag}(T)/T=0.711  (T/T_c)^2 +0.563 (T/T_c)^4-0.0628 (T/T_c)^6
\]

The self-energy decreases as the quark mass of the loop increases. The magnetic mass of a gluon through bottom quark loops is less than $k_BT_c$ and the thermal response of the bottom quark and the charmed quark differ qualitatively. 

When $T=0$, I evaluate the self-energy  as the 0  component of the Clifford product.
In the Coulomb gauge, the exchanged self-dual gluons are assumed to be $x_1,x_2$  and the quark-anti quark state that couple to $x_4$ can be  $\psi C\phi\to(\xi_1\xi_{23}+\xi_2\xi_{31})$, or
 $\phi C\psi\to (\xi_{14}\xi_{234}+\xi_{24}\xi_{314})$.  The intermediate quark-anti quark state independent of the momentum could be $\xi_{1234}\xi_4$ in the former case and $\xi_{123}\xi_0$ in the latter case.  However, since these states don't have virtual momentum excitations,  I consider, instead
transition of $\psi C\phi$ to  $\phi C\psi$ and vice versa in the intermediate state.

I define self-energy $\Pi_{44}^c$ for exchange of self-dual gluons  $x_1,x_2$ is obtained by using eigen states
$(\xi_{14}\xi_{234}+\xi_{24}\xi_{134} )$ or $ (\xi_{23}\xi 1+\xi_{13}\xi_2)$. 
\begin{eqnarray}
&&J_c(k,p)=m^4g^6 \frac{k_x}{m} Tr \gamma_1 \frac{-\gamma_1k_x+m}{k_x^2+m^2}\gamma_2\gamma_5\frac{1}{q_y^2}\frac{-\gamma_3 (k_x \times q_y)+m}{(k_x\times q_y)^2+m^2}\gamma_1\gamma_5\frac{1}{r_x^2}\frac{-\gamma_2 (r_x k_x) q_y+m}{(r_x k_x)^2 q_y^2+m^2}\nonumber\\
&&\times\frac{ p_y}{m} \gamma_2 \frac{\gamma_2 p_y+m}{p_y^2+m^2}\gamma_1\gamma_5\frac{\gamma_3 (p_y\times r_x)+m}{(p_y\times r_x)^2+m^2}\gamma_2\gamma_5\frac{\gamma_1 (q_y p_y)r_x+m}{(q_y p_y)^2 r_x^2+m^2}
\end{eqnarray}

\begin{figure}[htb]
\begin{center}
\includegraphics[width=6cm,angle=0,clip]{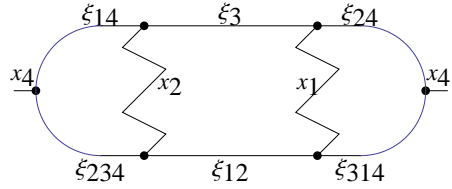}
\caption{The self-energy diagram of time-like gluon through exchange of self-dual gauge fields. 
$\Pi_{44}^c$}
\label{44a}
\end{center}
\end{figure}

When the exchanged self-dual gluons are $x_1 x_2$,  the trace in the numerator of $J_c(k,p)$  becomes
\begin{equation}
4m^6[({p_y}^2+{k_x}^2)q_y r_x-(1+(q_y r_x)^2)k_x p_y][m^2- (k_x\times q_y)( p_y\times r_x)]
\end{equation}
Numerically one could fix the momentum transfer $(k_x r_x)q_y-p_y$  and integrate over $r_x, k_x$ and $p_y$, or the momentum transfer $(q_y p_y)r_x-k_x$ and integrate over $q_y, p_y$ and $k_x$.

When the external gluon is a spacelike photon, the infrared divergence can be regularized and yields 
anomalous magnetic moment $\frac{g}{2}=1+\frac{\alpha}{2\pi}$.  
Since regularization of $q_y, r_x$ is necessary, the numerical calculation of the self-energy of the timelike gluons and photons remain as a future problem.
 
\section{The B-meson weak decay vertex }
When the two self-dual gluon exchange is important in the heavy quark heavy anti quark pair to a gluon coupling, similar two gluon exchange in B meson weak decay is expected to be important.
I consider $B^-\to \tau^-\bar\nu$ decay, in which $b$ quark $\bar u$ quark couple to $W$ boson through Cabibbo-Kobayashi-Maskawa matrix as,
\begin{eqnarray}
&&(\bar u(q+p),\bar c(q+p),\bar t(q+p))\gamma_\mu(1-\gamma_5) \left(\begin{array}{ccc} U_{ud}&U_{us}&U_{ub}\\
                                                                                  U_{cd}&U_{cs}&U_{cb}\\
                                                                                  U_{td}&U_{ts}&U_{tb}\end{array}\right)
\left(\begin{array}{c} d(q)\\
                             s(q)\\
                              b(q)\end{array}\right) W_\mu(p)^\dagger+hc\nonumber\\
&&=u(q+p)^* \gamma_4 \gamma_\mu(1-\gamma_5) U_{ub}\, b(q) \,W_\mu(p)^\dagger+\cdots
\end{eqnarray}
and its hermitian conjugates.

\begin{figure}[htb]
\begin{minipage}[b]{0.47\linewidth}
\begin{center}
\includegraphics[width=6cm,angle=0,clip]{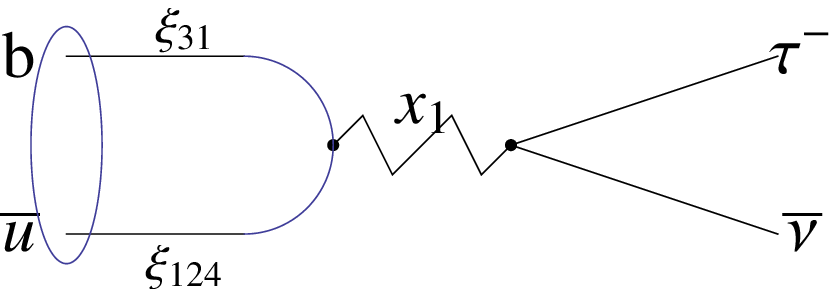}
\caption{The $B -\tau\bar\nu$ vertex. The direct term. $x_1$ is the transverse W boson.}
\label{vtx1}
\end{center}
\end{minipage}
\hfill
\begin{minipage}[b]{0.47\linewidth}
\begin{center}
\includegraphics[width=6cm,angle=0,clip]{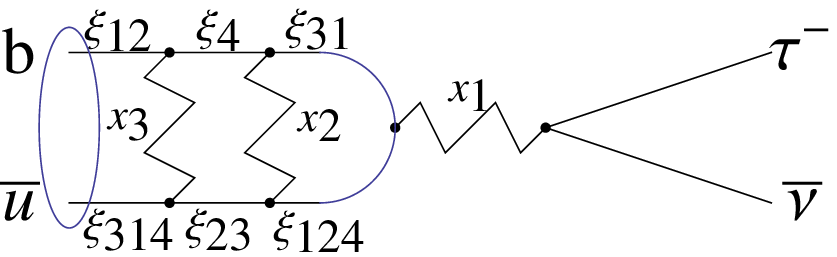}
\caption{The $B -\tau\bar\nu$ vertex. The three loop term. $x_1$ is the transverse W boson, $x_2$ and $x_3$ are self-dual gluons.}
\label{vtx2}
\end{center}
\end{minipage}
\end{figure}

The direct coupling of $W$ boson to the quark current has two types:
\begin{eqnarray}
D_a(q_y)&=& T g_W \iint Tr\frac{-\gamma_3  k_z+m_b}{k_z^2+m_b^2} \gamma_4\gamma_i (1-\gamma_5)\frac{\gamma_2 (q_y+p_y)+m_u}{(q_y+p_y)^2+m_u^2}\phi(q_y+p_y,k_z) \frac{dk_z dp_y}{(2\pi)^2} \nonumber\\
&=& T g_W \iint \frac{-4 k_z (q_y+p_y)}{(k_z^2+m_b^2) ((q_y+p_y)^2+m_u^2) } \,\phi(q_y+p_y,k_z)\frac{dp_y dk_z}{(2\pi)^2},
\end{eqnarray}
and
\begin{eqnarray}
D_b(q_z)&=& T g_W \iint Tr\frac{-\gamma_3  k_y+m_b}{k_y^2+m_b^2} \gamma_4\gamma_i (1-\gamma_5)\frac{\gamma_2 (q_z+p_z)+m_u}{(q_z+p_z)^2+m_u^2}\phi(q_z+p_z,k_y) \frac{dk_y dp_z}{(2\pi)^2} \nonumber\\
&=& T g_W \iint \frac{-4 k_y (q_z+p_z)}{(k_y^2+m_b^2) ((q_z+p_z)^2+m_u^2) }\,\phi(q_z+p_z,k_y)\frac{dp_z dk_y}{(2\pi)^2}
\end{eqnarray}
where $\phi(p,k)$ is the B meson wave function in momentum space normalized as
\[
\iint |\phi(p,k)|^2 dp dk=1.
\]

 At large $q$, the Fourier transform of Gaussian wave function $D(q)$ decreases too fast. Thus, I adopt the discrete cosine transform of the Bethe Salpeter equation evaluated by a lattice simulation\cite{MP98}. Details of the choice of $\phi(p,k)$ is given in the Appendix.  
 
The numerical value of $D_a(q_y)$ at $q_y=-\frac{M_B^2-m_\tau^2}{ 2 M_B} =-2.34$ GeV is $0.6  \mathcal N T g_W$, where $ \mathcal N=0.025$ is the wave function normalization. At $q_y=2.34$GeV, the magnitude is about $10^{-4}$ times smaller.

In the 3 loop diagram, I choose the quark momentum is 0 as the $b-$quark.
We are interested in the momentum region small compared to the temperature $T$, where the
thermal part of the gluon propagator is simplified as\cite{AN95}

\begin{equation}
\frac{\delta(k^2-M^2)}{e^{\omega_k/T}-1}\sim \frac{T}{2\omega_k^2}[\delta(k_0-\omega_k)+\delta(k_0+\omega_k)]
\end{equation} 
where $\omega_k=\sqrt{ k^2+M^2}$. 

The transition amplitude is obtained by calculating the trace,
\begin{eqnarray}
&&{C_a(T,q_z)}=T m_u^4   g_W g^4\iint
 Tr \frac{-\gamma_3  k_z+m_b}{k_z^2+m_b^2}\gamma_3\frac{T}{k_z^2+m_{mag}^2(T)}\frac{1}{m_b}\frac{T}{p_y^2+m_{mag}^2(T)}\gamma_2\frac{-\gamma_2 p_y+m_b }{p_y^2+m_b^2}\nonumber\\
&&\times(1-\gamma_5) \gamma_1\gamma_4 \frac{\gamma_3(q_z+ k_z)+m_u}{(q_z+k_z)^2+m_u^2} \frac{p_y\times k_z}{m_u} \gamma_2\gamma_5\frac{\gamma_2 p_y+\gamma_3( k_z+q_z)+m_u}{ (k_z+q_z)^2+p_y^2+m_u^2} \frac{p_y\times k_z}{m_u} \gamma_3\gamma_5  \nonumber\\
&&\times\frac{\gamma_2  p_y +m_u}{ p_y^2+m_u^2}\phi(p_y,k_z)
\frac{d p_y d k_z }{(2\pi)^{2} }\nonumber\\
&&=T^3 g_W g^4 { m_u}\iint \frac{4k_z^2 p_y^2(3m_b^2 m_u+m_u^3+k_z(q_z+k_z)(-m_b+m_u)+(m_b+m_u)p_y^2)}
{(k_z^2+m_b^2) ((q_z+k_z)^2+m_u^2) (p_y^2+m_b^2) ( p_y^2+m_u^2)((q_z+k_z)^2+p_y^2+m_u^2)}
\nonumber\\
&&\times\frac{ p_y (q_z+k_z)}{(k_z^2+m_{mag}^2(T)) (p_y^2+m_{mag}^2(T)) } \phi(p_y,k_z)\frac{dp_y dk_z }{(2\pi)^2}
\end{eqnarray}

\begin{eqnarray}
&&{C_b(T,q_x)}=T m_u^4 g_W g^4\iint 
 Tr  \frac{-\gamma_1 k_x+m_b }{k_x^2+m_b^2} \frac{T}{k_x^2+m_{mag}^2(T)}\frac{1}{m_b}\frac{T}{p_z^2+m_{mag}^2(T)}\gamma_3 \frac{-\gamma_3 p_z+m_b}{ p_z^2+m_b^2}\nonumber\\
&&\times(1-\gamma_5) \gamma_2\gamma_4  \frac{\gamma_1 (q_x+k_x)+m_u}{(q_x+k_x)^2+m_u^2} \frac{p_z\times k_x}{m_u} \gamma_3\gamma_5\frac {\gamma_3 p_z+\gamma_1( q_x+k_x)+m_u}{p_z^2+(q_x+k_x)^2+m_u^2} \frac{ p_z\times k_x}{m_u}  \gamma_1\gamma_5 \nonumber\\
&&\times  \frac{\gamma_3  p_z+m_u}{ p_z^2+m_u^2} \phi(p_z,k_x)\frac{d p_z d k_x}{(2\pi)^{2} }\nonumber\\
&&= T^3  g_Wg^4 {m_u} \iint \frac{4k_x^2 p_z^2(3m_b^2 m_u+m_u^3+k_x(q_x+k_x)(-m_b+m_u)+(m_b+m_u)p_z^2)}
{(k_x^2+m_b^2) ((q_x+k_x)^2+m_u^2) (p_z^2+m_b^2)  (p_z^2+m_u^2)(p_z^2+(q_x+k_x)^2+m_u^2) }\nonumber\\
&&\times\frac{ p_z (q_x+k_x)}{(k_x^2+m_{mag}^2(T)) (p_z^2+m_{mag}^2(T)) } \phi(p_z,k_x)\frac{dp_z dk_x}{(2\pi)^2}
\end{eqnarray}

When $T=T_c$, $\displaystyle q_z=-2.34$ GeV, and the numerical value of the integral $C_a(T,q_z)$  is $\displaystyle 2.75  \mathcal N T^3 m_u g_W (2\alpha(q_{eff}))^2 <  \mathcal N T^3  g_W$. 
Here, $g^2$ is evaluated  from phenomenological $2\pi (2 \alpha(q_{\rm{eff}}))$, which depends on the effective momentum transfer of self-dual gluons $x_2$ and $x_3$. I assumed $\alpha(q_{\rm{eff}})\leq \pi$.  

In the case of B meson decay, the momentum transfer $q$ is of the order of $M_B/2$, and the
damping of $\phi(p+q,k)$ etc are important, and for an exact evaluation, lattice simulations are necessary.

There is no infrared divergence even when $m_{mag}(T)=0$, and the finite $m_{mag}(T)$ makes the vertex suppression at high $T$.

The decay rate $\displaystyle\Gamma=\frac{1}{\tau}$ of $B\to l\bar\nu$ in PQCD is\cite{Hou93,BK99}
\begin{equation}
\Gamma(B\to l\bar\nu)=\frac{1}{8\pi}|U_{ub}|^2G_F^2 f_{B}^2 M_B^3| \mathcal M |^2(1-\frac{m_l^2}{M_B^2})^2
\end{equation}
where $\langle0|\bar u\gamma_\mu \gamma_5 b|B^-\rangle=if_B p^\mu_B$ and
 $\mathcal M=\frac{m_l}{M_B} \bar l(1-\gamma_5)\bar\nu$.  Including charged Higgs effect, $f_B|U_{ub}|\sim 0.85$ is reported\cite{Hou93}.

\begin{equation}
\Gamma(B\to l\bar\nu)=\frac{1}{8\pi}|U_{ub}|^2 M_B| D(T) +C(T) |^2(1-\frac{m_l^2}{M_B^2})^2
\end{equation}

The correction $C(T)$ to $D(T)$ is not too large in the case of $B_u=\bar b u$ meson. 
Experimentally, $\tau(B^+)\geq \tau(B^0)>>\tau({B_c}^+)$.  A Higgs enhancement effect
in $b\to \tau\bar\nu+X$ \cite{Hou93} and R-parity violation in $B\to l\bar\nu$ \cite{BK99}
are discussed with an estimation $f_{B_c}=450$MeV\cite{BK99}.

\section{Discussion and Conclusion}
In this paper, an $\omega$ independent, temperature dependent magnetic mass of a gluon is calculated in a model with triality symmetry.  I checked the consistency of the model by estimating a correction to the width of a B meson decay into a lepton and an anti-neutrino. The idea of triality symmetry comes from the difference of the critical flavor number for opening the conformal window in the Schr\"odinger functional method\cite{AFN08} and in the Domain Wall Fermion method\cite{SF09}.
The polarized electron experiment at JLab\cite{JLab07} and the QCD running coupling constant in holographic QCD\cite{BTD10} and in the Coulomb gauge lattice simulation in MOM scheme\cite{FN07b}
suggest that the flavor number 2+1 system is not far from the conformal window, but the Schr\"odinger functional method suggests that the flavor number should be more than 8 for the conformal window opens. If DWF or MOM scheme selects one triality sector, while the Schr\"odinger functional does not due to the difference of the boundary condition of the wave function, the difference of about a factor of three in the critical number of flavors could be understood.

The non-perturbative effects in finite temperature QCD is difficult.  In \cite{TV80}, problems in renormalization of the three loop diagram as $J(p,k)$ or $C(p,k)$ are discussed.  If I replace the propagator denoted by $\xi_4$ is replaced by a point, an integral like
\[
J=\iiint \frac{dp\, dq\, dt(qt)^2}{p^2 q^2 t^2 (p-q)^2(k-q)^2 (k-t)^2}
\]   
appears. The momenta of internally exchanged gluons $p$ and $k$ in my model correspond to their $t$ and $q$.  The momenta corresponding to $\xi_{23}$ or $\xi_{31}$ in Fig.\ref{g622a} or Fig.\ref{11a} in \cite{TV80} are not fixed, since a coupling of the external gluon to the quark $\xi_4$ is considered in their model. Their $p$ is taken as an integration variable, but this complication is absent in our model. 

I approximated the quark wave function by the Bethe Salpeter wave function derived in a lattice simulation\cite{MP98}. It would be possible to refine the wave function and extend to the weak decay of $B_c$, which is discussed in \cite{BK99} as a check of the R-parity violation in the Minimal Supersymmetric Standard Model(MSSM). How the vertex correction of the present model affects the  charged Higgs Boson effects in the weak decay of B meson\cite{Hou93} is left to the future.  

In the model of domain wall fermion, the quark spinor possesses an arbitrary $U(1)$ phase that one can adjust to one triality sector of the lepton in the detector. I expect that there is three fold $U(1)$ symmetry in the quark sector, but the symmetry is broken in the lepton sector and one  triality is selected.  The situation seen from the side of a quark is similar to the color-flavor locking\cite{Wilczek11}.  In color-flavor locked states, the photon coupling to quark flavor $\gamma$ and gluon coupling to quark color $\Gamma$ have the same structure
\[
 \gamma/\Gamma=\left(\begin{array}{ccc}2/3& 0 & 0\\
                                           0 & -1/3 & 0\\
                                           0 & 0 & -1/3\end{array}\right)
\]
and the combination $\tilde\gamma=\frac{g\gamma+e\Gamma}{\sqrt{g^2+e^2}}$ defines a massless gauge boson. 

This hypothesis might explain the origin of dark matter or the weakly interacting massive particles (WIMPs), which are not detected by electro-magnetic detectors on the earth. If the vector field of a photon can also be expressed by using Pl\"ucker coordinates, it could also acquire a magnetic mass.

 \vskip 0.3 true cm
\leftline{\bf Acknowledgement}
I thank Prof. Stanley Brodsky for very helpful suggstions and encouragements and Prof John Cornwall for his kind interest on the magnetic mass.
The support of using super computers in KEK, YITP of Kyoto Univerity and in Tsukuba University are  thankfully acknowledged. 
\newpage
\leftline{\Large{\bf Appendix: Quark wave function $\phi(p,k)$}}

The discrete cosine transform(DCT) is used in image compressions etc. For a given data  $s(x)$ ($x=0,1,\cdots N-1$) one multiplies $\displaystyle \cos \frac{\pi k(2x+1)}{2N}$ and make a weighted summation\cite{Wa94,SF07}. When the DCT is applied to $\displaystyle \sqrt{\frac{\beta}{\pi}} e^{-\beta x^2/2}$, an expresion close to its Fourier transform $\displaystyle \sqrt{\frac{1}{\pi\beta}} e^{-p^2/2\beta}$ is obtained. 

\begin{figure}[htb]
\begin{minipage}[b]{0.47\linewidth}
\begin{center}
\includegraphics[width=6cm,angle=0,clip]{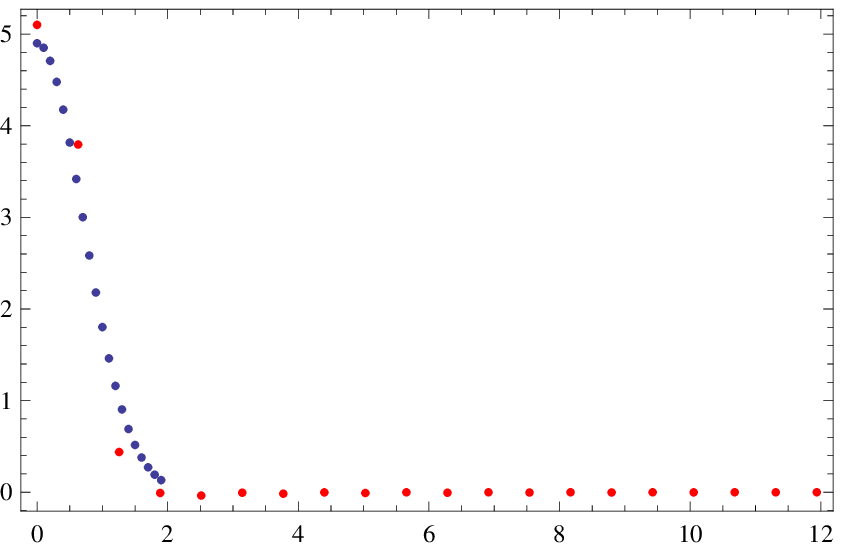}
\caption{The discretized gaussian wave function and its DCT.}
\label{dctgauss2}
\end{center}
\end{minipage}
\hfill
\begin{minipage}[b]{0.47\linewidth}
\begin{center}
\includegraphics[width=6cm,angle=0,clip]{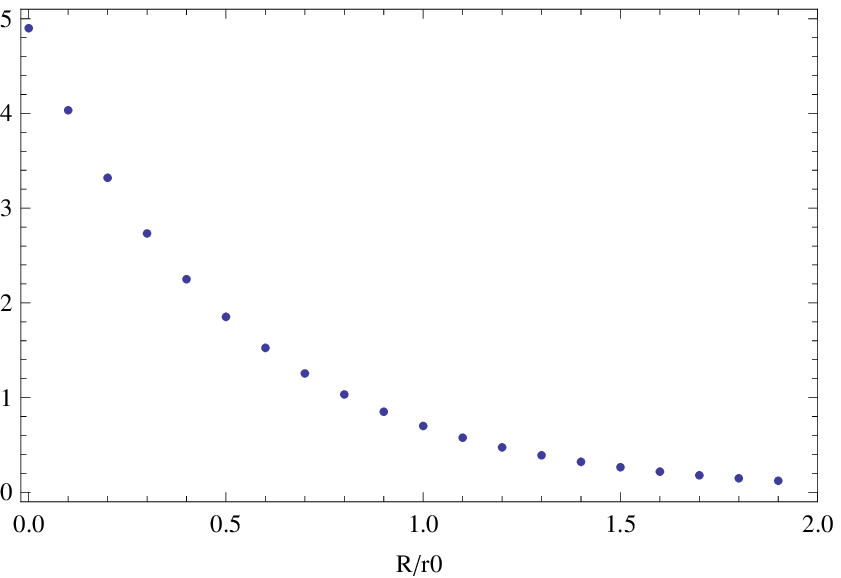}
\caption{The discretized Bethe Salpeter wave function.}
\label{dctw}
\end{center}
\end{minipage}
\end{figure}

I consider the case $\beta=2$ and evaluate DCT of $4.9 e^{-x^2}$. The factor 4.9 is the normalization that is used in the Bethe Salpeter wave function of the B meson which will be discussed later.
When the DCT is applied, the scale of ordinate changes by the factor $1/\beta$ and that of abscissa changes by $2\pi$. The DCT scaled by multipying $1/2$ is compared with the original in Fig.\ref{dctgauss2}. When the maximum of $x=2$ corresponds to 1fm, $\Delta x=0.05$ and in momentum space $\Delta p={2\pi}$ fm$^{-1}$=1.18 GeV. 

The parameter of the Gaussian wave function is taken from the flux-tube model\cite{IP85, KI87}. The model was applied to low energy meson decays and meson-baryon couplings and produced reasonable results. 

The Bethe Salpeter wave function of the $\bar Q q$ meson state was calculated in \cite{MP98}.  They adopted the scale $r_0=3a=2.68$GeV$^{-1}$=0.529fm, and the s-wave $w(R)$ can be parameterized as
\[
w\left(\frac{R}{r_0}\right)=4.9 e^{-\log 7 \frac{R}{r_0}}=4.9\times 7^{-\frac{R}{r_0}} .
\]
The wave function is normalized as $\displaystyle \int_0^\infty w\left(\frac{R}{r_0}\right)^2\left(\frac{R}{r_0}\right)^2 \frac {dR}{r_0}\sim 1$.

\begin{figure}[htb]
\begin{center}
\includegraphics[width=6cm,angle=0,clip]{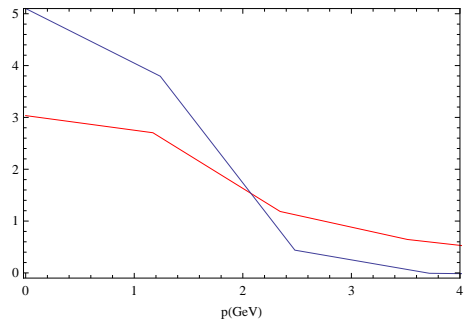}
\caption{The DCT of the discretized Bethe Salpeter wave function and the DCT of the Gaussian wave function.}
\label{dctMP2}
\end{center}
\end{figure}

I take 20 points of $\frac{R}{r_0}$ from 0 to 1.9, which corresponds to radial distance between $b$ anti quark and $u$ quark of about 1fm.  I make a discrete cosine transform of this wave function
shown in Fig.\ref{dctw}.
I multiply the scale of the ordinate $1/2$ at a moment.  Since $\Delta P$ corresponds to $2\pi /(2\times 0.5287){\rm fm}^{-1}=5.94 {\rm fm}^{-1}=1.17$GeV, I change the scale of abscissa and plot in Fig.\ref{dctMP2}.
The wave function $\tilde w(k)$ obtained by DCT is normalized as
\[
\sum_{j=0}^{19}  \tilde w(j)^2 (\frac{j+1}{10})^2 \sim 1
\]
When the scale of the ordinate is multiplied by $\frac{1}{2}$ as in the case of Gaussian, and abscissa is transformed to GeV, 
\[
\sum_{j=0}^{19} \frac{1}{4} \tilde w(j)^2 4\pi (1.17 ({j+1)}{\rm GeV})^2 \times  \Delta P\sim \frac{1}{{\mathcal N}^2}
\]
Since the Gaussian wave function can be decomposed as $\displaystyle e^{-(p^2+k^2)/(2\beta)}=e^{-[(\frac{p-k}{\sqrt 2})^2-(\frac{p+k}{\sqrt 2})^2]/(2\beta)}$ and the Bethe Salpeter wave function corresponds to
the $\displaystyle e^{-(\frac{p-k}{\sqrt 2})^2/(2\beta)}$ part, I identify $\displaystyle \frac{\mathcal N}{2}\tilde w(j)$
 as $\displaystyle \phi(\frac{p-k}{\sqrt 2})$ and the wave function of the center of mass to be 1, i.e. I regard $b$ quark is infinitely heavy.

The wavefunction $\phi(p,k)$ is approximated as a function of $|\Vec p-\Vec k|/\sqrt 2$. 
\begin{table}
\begin{center}
\begin{tabular}{ccccccc}
\hline
$p$ & 0 & $2\pi$ & $4\pi$ & $6\pi$ & $8\pi$ & fm$^{-1}$\\
$\phi_{gauss}$ & 5.10 & 3.80 & 0.44 & -0.01 &-0.04 &\\
\hline
$p$ & 0 & $5.94$ & $11.9$ & $17.8$ & $23.8$& fm$^{-1}$\\ 
$\phi_{MP}$ & 3.03 & 2.70 & 1.19 &0.65 & 0.37& \\
\hline
\end{tabular}
\caption{The momentum dependence of the relative wavefunction of quarks.}
\end{center}
\end{table}

\end{document}